\documentclass[prb,aps,showpacs,twocolumn,superscriptaddress,longbibliography]{revtex4-1}
\usepackage{graphics, bm, xspace, color}
\usepackage{psfrag}
\usepackage{amsmath}
\usepackage{amssymb}
\usepackage{epsfig}
\usepackage{subfigure}
\usepackage{grffile}
\usepackage{afterpage}
\usepackage[colorlinks=true,urlcolor=blue,linkcolor=blue,citecolor=blue]{hyperref}
\newcommand{\beq}    {\begin{equation}}
\newcommand{\enq}    {\end{equation}}
\newcommand{\ceq}[1] {(\ref{#1})}

\newcommand{\df}     {\equiv}

\newcommand{\dav}    {\langle D\rangle}
\newcommand{\tav}    {\langle T_c\rangle}

\newcommand{\rins}   {R_{I}}
\newcommand{\rmetal} {R_{M}}

\newcommand{\vo}     {${\rm VO_2}$\xspace}
\newcommand{\tio}    {${\rm TiO_2}$\xspace}
\newcommand{\alo}    {${\rm Al_2O_3}$\xspace}
\newcommand{\vosap}  {${\rm VO_2/Al_2O_3}$\xspace}
\newcommand{\votio}  {${\rm VO_2/TiO_2}$\xspace}

\graphicspath{{../Figures/}}

\begin{document}

\title{Effect of inhomogeneities and substrate on the dynamics of the metal-insulator transition in \vo thin films}
\author{M. Rodriguez-Vega}
\affiliation{Department of Physics \ College of William and Mary \ Williamsburg VA, 23187}
\author{M. T. Simons}
\affiliation{Department of Physics \ College of William and Mary \ Williamsburg VA, 23187}
\author{E. Radue}
\affiliation{Department of Physics \ College of William and Mary \ Williamsburg VA, 23187}
\author{S. Kittiwatanakul}
\affiliation{Department of Material Science \ University of Virginia \ Charlottesville VA, 22904}
\author{J. Lu}
\affiliation{Department of Material Science \ University of Virginia \ Charlottesville VA, 22904}
\author{S. A. Wolf}
\affiliation{Department of Physics \ University of Virginia \ Charlottesville VA, 22904}
\affiliation{Department of Material Science \ University of Virginia \ Charlottesville VA, 22904}
\author{R. A. Lukaszew}
\affiliation{Department of Physics \ College of William and Mary \ Williamsburg VA, 23187}
\author{I. Novikova}
\affiliation{Department of Physics \ College of William and Mary \ Williamsburg VA, 23187}
\author{E. Rossi}
\affiliation{Department of Physics \ College of William and Mary \ Williamsburg VA, 23187}
\date{\today}

   
\begin{abstract}
We study the thermal relaxation dynamics of  \vo films 
after the ultrafast photo-induced metal-insulator transition for two  \vo film samples grown on  \alo and \tio substrates. We find two orders of magnitude difference in the recovery time (a few ns for the \vosap sample vs. hundreds of ns for the\votio sample). 
We present a theoretical model to take into account the effect of inhomogeneities in the films
on the relaxation dynamics. 
%
We obtain quantitative results
that show how the microstructure of the \vo film and the thermal conductivity
of the interface between the \vo film and the substrate 
affect long time-scale recovery dynamics. 
We also obtain a simple analytic relationship between the recovery time-scale 
and the film's parameters. 

\end{abstract}


\maketitle


\section{Introduction}

Vanadium dioxide (\vo) undergoes a metal-insulator transition (MIT) around room temperature
\cite{morin1959} enabling a wide range of potential applications. 
It has recently been 
shown that it is possible to photo-induce the insulator-to-metal transition in \vo in the sub-picosecond timescale~\cite{Cavalleri1999d,Cavalleri2001a,Cavalleri2004a,Cavalleri2005a,Kim2006,Nakajima2009a,Cocker2010}. This finding makes \vo a material of great interest for electronic and photonic applications, such as ultra-fast switches or transistors. The realization of \vo-based switches requires the ability to control the \vo MIT dynamics using external fields, as well as a better understanding of the recovery mechanisms after the external field 
is turned off and the material returns to its normal state.
The mechanism by which the photo-induced insulator-to-metal
takes place in \vo is still not fully understood due to the complexity of
the electronic behavior of \vo arising from the presence of strong
electron-lattice coupling and electron-electron interactions 
\cite{goodenough1971,zylbersztejn1975,pouget1975,eyert2002}.
As a result, \vo is a unique material of great fundamental
and practical interest.

At low temperatures ($T\lesssim 340$~K) the \vo lattice has a monoclinic structure, whereas 
at high temperatures ($T\gtrsim 340$~K) it has a tetragonal structure.
This difference in lattice structure is reflected in the band structure:
\vo is an insulator in the monoclinic phase and a metal in the tetragonal phase.
This simple picture is complicated by the fact that in \vo electron-electron
correlations are very strong and can provide an important contribution
to the localization of the electronic states via the Mott mechanism \cite{zylbersztejn1975, pouget1975, paquet1980, stefanovich2000}.
It appears that a full account of the MIT must take into account
the interplay of the lattice dynamics and the electron dynamics driven
by strong electron-electron interactions. This is a fascinating and
extremely challenging problem that in addition is complicated
by the unavoidable presence of inhomogeneities 
\cite{qazilbash2007,wang2013}.

Several works  \cite{becker1994b,Kim2006,kubler2007, hilton2007,rini2008, pashkin2011, abreu2012,cocker2012, wall2013,radue2015}
have investigated the short timescale dynamics after the photo-induced transition.
In particular, Ref. \onlinecite{lysenko2007} presented a comparison of the long timescale recovery dynamics 
between \vo films on a crystal substrate or a glass substrate and found that the
recovery time for the films on the glass substrate was much longer than for the films
on a crystal substrate. The recovery time was modeled using the heat equation 
to describe the heat flow across the interface between the \vo film and the substrate.
The difference in the characteristic time between the two types of substrates
was attributed to the fact that the thermal conductivity of the interface
was expected to be much smaller for glass substrates than for crystal substrates.

In this work we present a theory to properly take into account the effect of inhomogeneities
on the recovery dynamics of \vo films. Our theory 
describes simultaneously:
(i)   the profile of the reflectivity across a thermal induced MIT;
(ii)  the long timescale recovery dynamics of the reflectivity after a photo-induced insulator-to-metal transition;
(iii) the observed difference of two orders of magnitude between samples with different substrates.
Inhomogeneities are due to the fact that the film is comprised
of grains with different sizes and different local properties, such as strain \cite{brassardAPL05,alievPSS06}  and chemical composition.

\begin{figure}[hbt]
    \includegraphics[width=8.5cm]{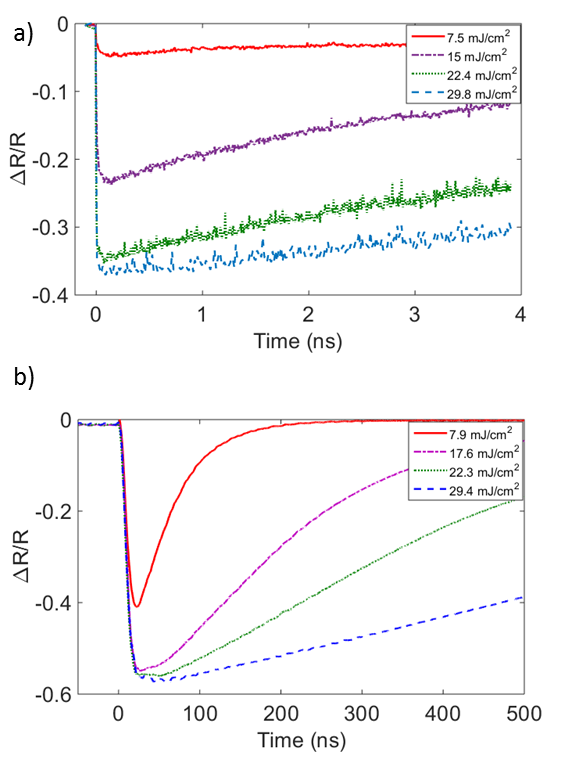}
    \caption{Relative change in reflectivity ($\Delta R/R$) for the \vo film on (a) \alo substrate and (b) \tio substrate as a function of time after the MIT is induced at time $t=0$ by a strong ultrafast pump pulse. The values of the pump fluence are shown in the legend, and the sample temperature is set to 311~K in (a) and 280~K in (b), which correspond to approximately $30$~K below the critical temperature $T_c$ for thermally-induced MIT for each sample.}
   \label{fig:04}
\end{figure}

The presence of inhomogeneities induces a distribution of values for the transition
temperature $T_c$ within the film. To take this into account
we derive a generalized heat equation that includes the fact
that during the recovery from the photo-induced insulator-to-metal transition,
at any given time a fraction of the sample is undergoing the metal-to-insulator
transition, another fraction is still cooling in the metallic phase,
and another fraction is already cooling in the insulating phase.
A key ingredient of the generalized heat equation is the correct description of the time evolution
of the fraction of the sample that is metallic, insulating, or undergoing the phase transition.
We then use our theoretical model to obtain the scaling relation between
the characteristic recovery time $\tau$ and the parameters of the films.
Our theoretical model, and the underlying assumptions, are strongly supported
by our experimental results. Differently than in Ref.~\onlinecite{lysenko2007}
our \vo films have all crystal substrates, no glass. Yet, we find that
$\tau$ can be more than two orders of magnitude different depending on the crystal
substrate, \tio or \alo, Fig.~\ref{fig:04}.

The generalized heat equation,
Eqs.~\ceq{eq:dQ2}, which properly takes into account the effect of
the films inhomogeneities on the recovery dynamics is the main result of our work.
Our theory allows the description of the recovery dynamics consistently
with the measurements obtained for the thermally driven MIT.
The scaling between the characteristic 
recovery time $\tau$ and the parameters of the film is another important result of our work.

Our work is relevant to the more general problem of how spatial 
inhomogeneities affect a first order phase transition.
The ability of our treatment to contribute to this general problem relies
on the fact that in \vo the two phases across the first-order phase transition
have very different electronic properties (metallic vs. insulating behavior)
that allows us to get an accurate phase mapping, via optical reflectivity measurement,
of the time evolution of the metallic (insulating) fraction and,
indirectly, of the spatial inhomogeneities present during the transition.

The work is organized as follows.
Sec.~\ref{sec:exp} describes the experimental arrangements to measure the optical reflectivity time-evolution. The details of the theoretical model that we use  to characterize
the distribution of the films' inhomogeneities and the 
long-time dynamics of the reflectivity after a photo-induced insulator-to-metal transition are presented 
in Secs.~\ref{sec:th_model_static} and ~\ref{sec:th_model_dynamics}, respectively. In Sec.~\ref{sec:results} we demonstrate how the variations in statistical properties of the two films result in a significant difference in the relaxation timescales, and 
in Sec.~\ref{sec:conclusions} we provide our conclusions.



\section{Experimental setup}
\label{sec:exp}

In our experiments we studied two VO$_2$ thin-film samples, both of which were produced using 
reactive-bias target ion beam deposition (RBTIBD)~\cite{West2008a}. 
One sample was grown on $0.5~$mm thick c-Al$_2$O$_3$, and the thickness of the VO$_2$ film was 80 nm. The other sample was grown on a $0.5$ mm thick TiO$_2$ (011) substrate, and was measured to be 110 nm thick. X-ray diffraction (XRD) evaluation of both films showed them to be crystalline, and detailed characterization information is available in previous reports~\cite{Wang2012d,radue2013}. 

\begin{figure}[!ht]
    \includegraphics[width=8.0cm]{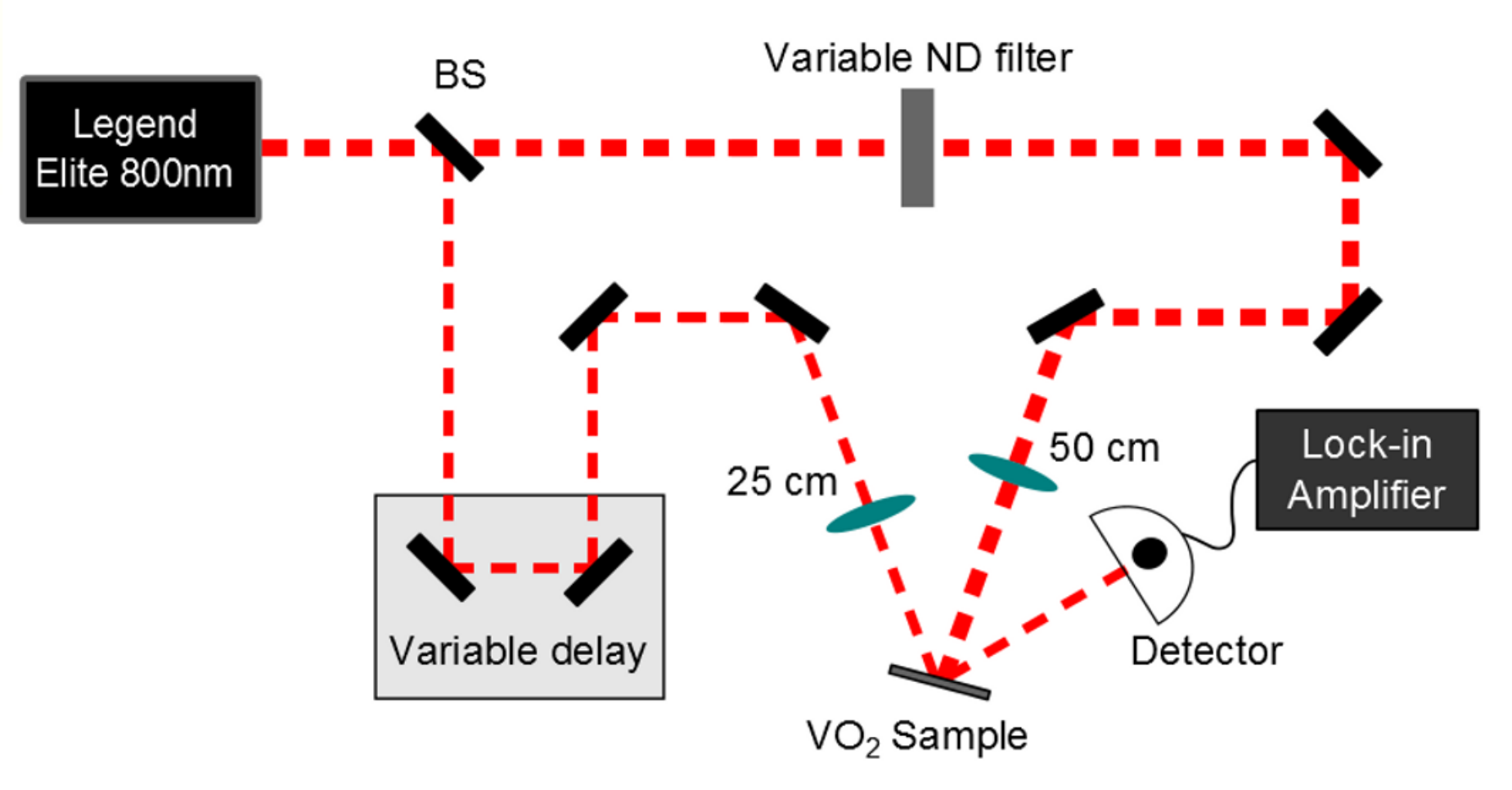}
    \caption{Schematic of the ultrafast pump-probe setup. BS is an $80/20$ beam splitter.}
    \label{setup_fast}
\end{figure}

For the photo-induced insulator-to-metal transition experiments we used an ultrafast laser system (Coherent Mantis oscillator and Legend Elite regenerative amplifier) with approximately $100$~fs pulses with a central wavelength at  $800$~nm and a repetition rate of 1~kHz.
The properly attenuated output of the laser was split into strong pump pulses and weaker probe pulses using a beam splitter in a standard pump-probe configuration, shown in Fig.~\ref{setup_fast}. The more powerful pump beam, focused to a $180~\mu$m diameter spot on the surface of the sample, was used to induce the insulator-to-metal transition, and its fluence was controlled using a variable neutral-density filter (VF). The fluence of the probe beam was further attenuated to a value well below the insulator-to-metal threshold ($\phi_\mathrm{probe} \le 100~\mu$J/cm$^2$), and we used its reflectivity from the sample to monitor the instantaneous optical properties of the VO$_2$ film.  The probe pulses were directed along a variable delay stage to accurately control the relative timing between the pump and probe pulses by up to $4~$ns with a few fs precision. The probe beam was focused on the sample at the same spot as the pump beam, using a shorter focal length lens. When tuned to the center of the pump beam focal spot, the smaller probe beam diameter ($90~\mu$m) ensured probing a region of  uniform pump intensity.

The reflected probe power was measured using a silicon photodetector, and further analyzed using a lock-in amplifier. To minimize the effects of probe pulse instabilities, as well as long-terms drifts due to environmental changes, we report the relative change in probe reflection $\Delta R/R$ with the pump beam on or off.

\begin{figure}[!!!!h!!!!!bt]
    \includegraphics[width=7.0cm]{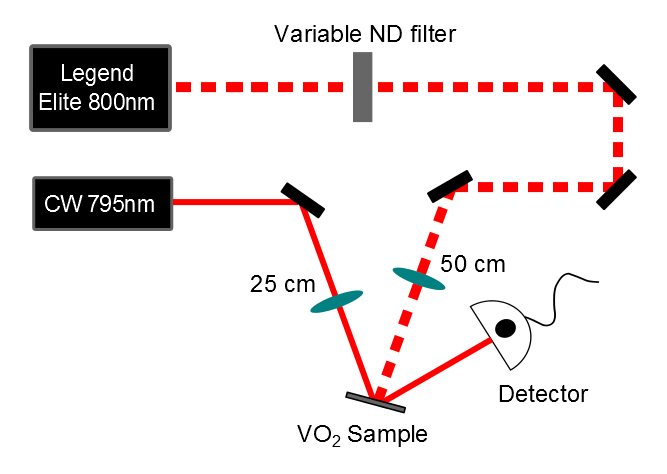} 
    \caption{Schematic of the experimental setup using a continuous-wave probe laser.}
    \label{setup_slow}
\end{figure}

Notably the MIT relaxation of the \vo/\tio  sample was not measurable with the femtosecond probe, as its characteristic decay time exceeded the $4~$ns maximum pulse separation, determined by the length of the delay stage. To measure the relaxation of the metallic VO$_2$ grown on the rutile sample we modified our experimental setup by replacing the femtosecond probe pulses with a continuous-wave (CW) diode laser operating at $785$~nm  and a fast photodiode (measured response time of approximately $10$~ns), as shown in Fig.~\ref{setup_slow}. 
This detection method allowed us to measure changes in reflectivity for times longer than $\approx 20$~ns  after the insulator-to-metal transition, that were inaccessible with the femtosecond probe arrangement. 
\begin{figure}[!!!!h!!!!!bt]
    \includegraphics[width=8.5cm]{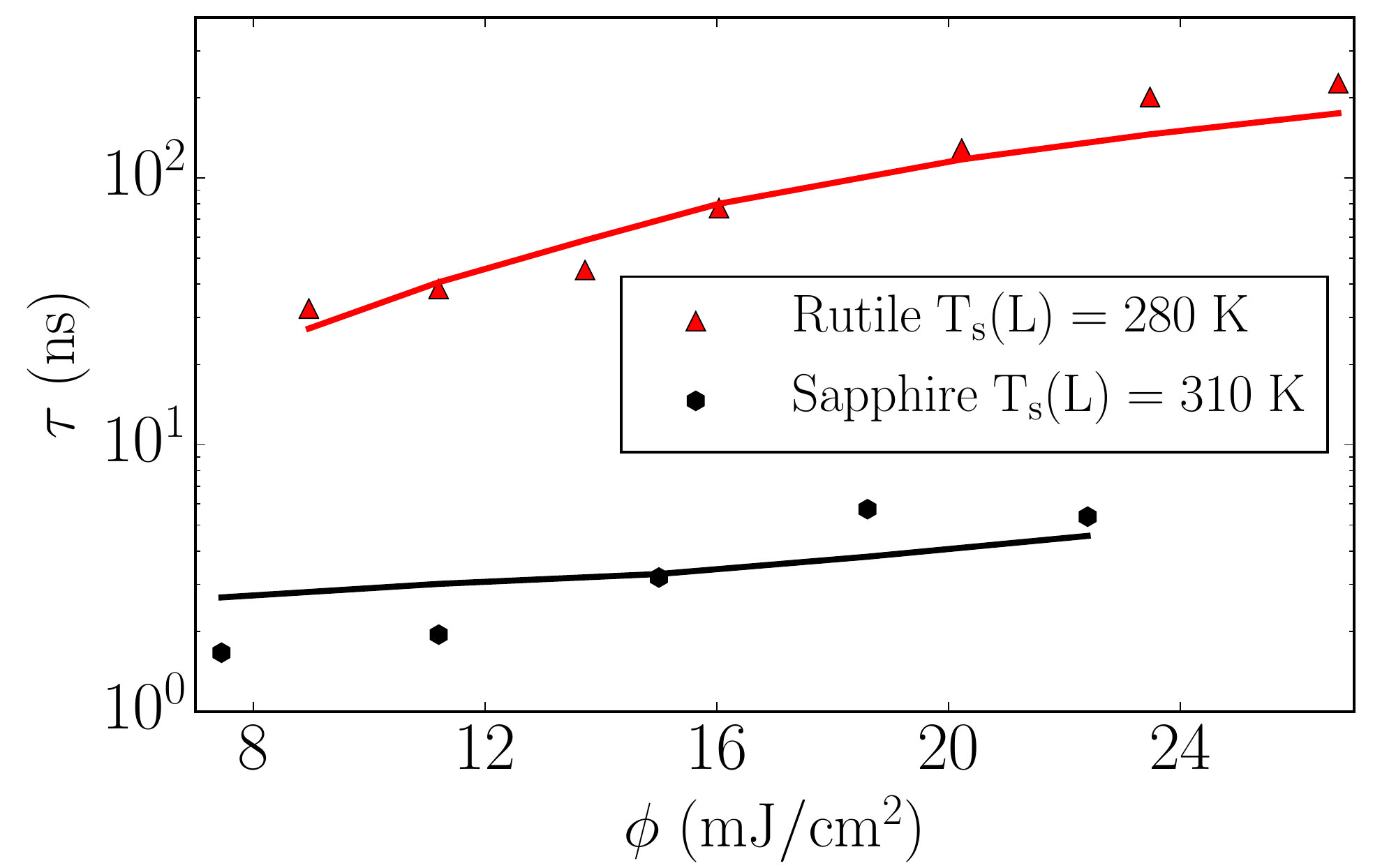}
    \caption{Dependence of metal state decay constant $\tau$ on the laser pump fluence and substrate temperature. Dots represent experimental data, and lines corresponds to the results of the theoretical calculations. The initial temperature $T_s$ for both sample substrates was approximately $30$~K below their respective MIT critical temperatures. }
    \label{fig:09}
\end{figure}
%

%
Figure~\ref{fig:04} shows sample measurements of both the \vosap and \votio films, using the femtosecond and CW probe arrangements respectively.
The overall reflectivity depends on the refractive index of both the film and the substrate, and the refractive indices of \tio and  \alo are different. 
Because it is easier to average the CW laser reflection signal, the curves for \votio are smoother than the curves for the \vosap.
The rutile reflection spectra recorded using the ultrafast probe had the same noise as for the sapphire samples, indicating that the 
differences in the noise are due to differences in the probes, not in the samples.
%

For values of the pump fluence higher than a threshold, which depends on the substrate temperature, 
we can see that the reflectivity, soon after the pump pulse, remains almost constant
for some time, i.e. its dynamics exhibits a ``flat'' region, see in particular Fig.~\ref{fig:04}~(b). 
The observed ``flattening'' of the curves is due to the pump pulse heating the sample to a temperature above the threshold 
value for the thermally-induced insulator-to-metal transition \cite{cocker2012,radue2015}. 
In this case the reflectivity stays unchanged at the level corresponding to a fully metallic phase
until a non-negligible fraction of the sample cools
down to the transition temperature. For all experimental curves only the later exponential part of the measured reflectivity was included into the fitting thermal relaxation time analysis. 
%
%
 
%
The analysis of the relative reflectivity for both \vo samples demonstrate that after the initial rapid change during the ultra-fast 
insulator-to-metal transition, its time evolution during the recovery is well fitted by a single exponential function with a recovery time constant $\tau$: 
%
%
\begin{equation}
R_{\rm fit}(t)=R_{I} + \left( R_0 - R_{I} \right)e^{-t/\tau},
\end{equation}
where $R_{I}$ corresponds to the reflectivity in the insulating phase, and $R_0$ corresponds to the reflectivity at $t=0$~s. The results of such measurements are shown in Fig.~\ref{fig:09}: for \vosap films we obtained values of $\tau$ of the order of few nanoseconds, whereas
it took the \votio sample a few hundred nanoseconds to relax back to the insulating state. This two orders of magnitude difference in the recovery times was even more surprising considering that the characteristic times for the transition itself were quite similar, as demonstrated in previous studies~\cite{radue2015}. In the discussion below we demonstrate that the relaxation dynamics strongly depend on the microstructure of the \vo films which in turn is strongly influenced by the properties of the substrate and their interface.
Figure~\ref{fig:09} also reveals that the rate of thermal relaxation for both samples increases with higher pump power. 
%
%


\section{Theoretical modeling of inhomogeneities}
\label{sec:th_model_static}
\begin{figure}[!!!!h!!!!!bt]
    \includegraphics[width=8.5cm]{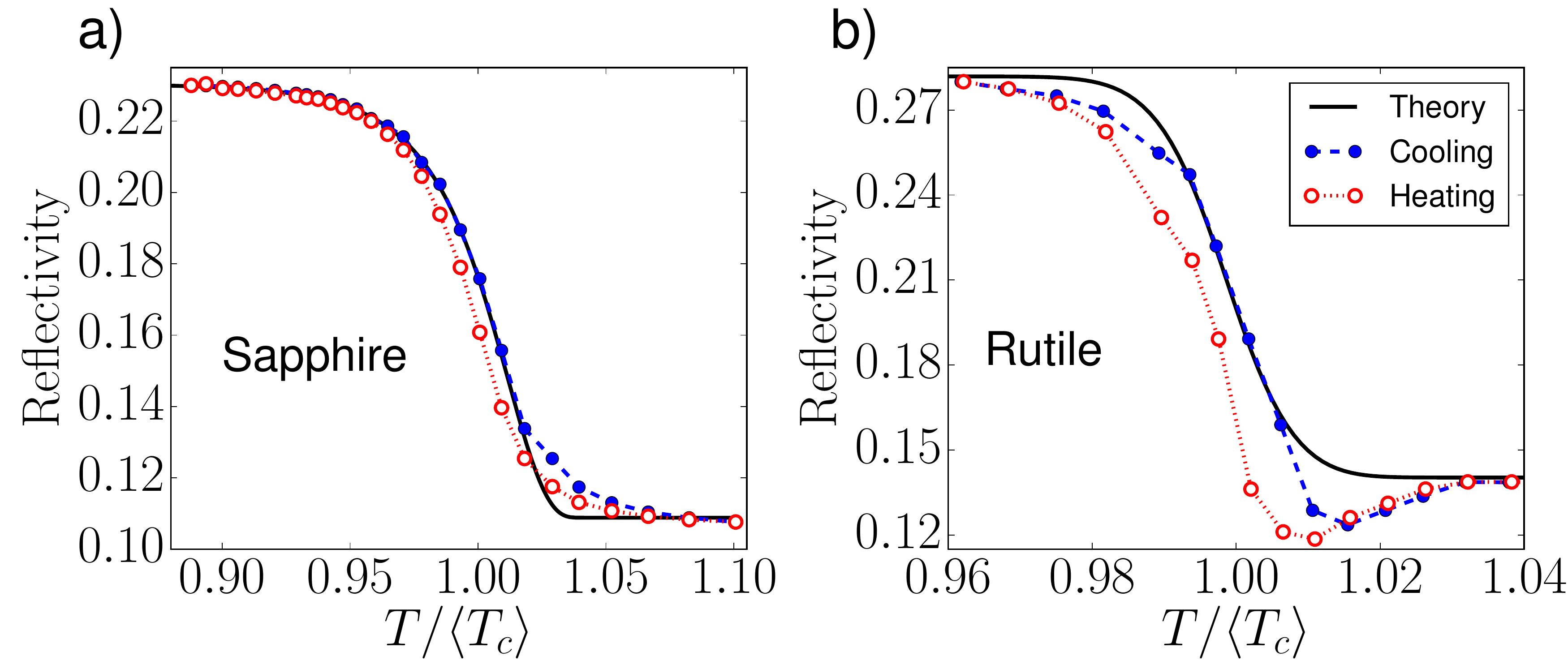}
    \caption{Evolution of the reflectivity across the thermally induced MIT for the case of sapphire and rutile substrates normalized to the average critical transition temperature. The open circles (red) correspond to the measured reflectivity in the heating branch, the solid circles (blue) correspond to the measured reflectivity in the cooling branch, and the solid curve corresponds to the theoretical result.
    For rutile substrate $\langle T_c \rangle = 314.0$ K, and for the sapphire substrate $\langle T_c \rangle = 340.1$ K. } %
   \label{fig:01}
\end{figure}

In order to take into account the effect of the inhomogeneities on the MIT dynamics
the first step is to characterize them.
%
%
To do this we can use the profile of the reflectivity across the thermally induced MIT.
The dotted lines in Figures~\ref{fig:01}~(a), and (b) show the measured reflectivity 
as a function of temperature across the thermally induced MIT for a \vo film grown on sapphire and \tio, respectively.
The temperature driven MIT in \vo is a first-order transition. 
In the ideal case
the reflectivity is expected to exhibit a finite, step-like, change at the critical temperature $T_c$, at which the sample goes from a low-temperature insulating state to a high-temperature metallic state. In thin films, however,  
the optical reflectivity smoothly changes from the value corresponding to the insulating phase ($\rins$) to the value characteristic to the metallic phase ($\rmetal$) as the temperature increases, as Fig.~\ref{fig:01} illustrates. 
For our samples the hysteresis loop is very narrow \cite{radue2013}.
The fact that the MIT takes place over a range of temperatures implies that different regions of 
the sample have different values of $T_c$.
This is different from the case of an ideal, homogeneous, system for which the whole sample exhibits 
the coexistence of metallic and
insulating regions only for $T=T_c$.
As a consequence the MIT in the films is characterized not by a single critical temperature but
by a distribution $P(T_c)$ of critical temperatures. This is due to the fact that
the \vo films are inhomogeneous: they are formed by crystal grains with different local properties.
Different grains in general have different sizes, slightly different stoichiometry, and experience
different local strains. It is very challenging to characterize the distribution of all the local
properties that can affect the transition temperature of each grain. 
However, for our purposes, we only need $P(T_c)$ and, as we show below, this can obtained directly from
the profiles of $R(T)$ without having to characterize the distribution of the local
properties affecting $T_c$.
Let $\eta_I$ be the fraction of the sample in the insulating phase. 
At a given temperature $T$ we have:
\beq
 \eta_I(T)=\int_T^\infty P(T_c)dT_c.
 \label{eq:etaI}
\enq
Let $\eta_m(T)=1-\eta_I(T)$ be the fraction of the film in the metallic phase.
To obtain the evolution of $\eta_I(T)$ across the MIT, and therefore $P(T_c)$,
considering that changes in the fraction of the film that is metallic (insulating) are the dominant
cause of changes in the reflectivity, we can use 
a two-fluid effective medium theory (EMT)
\cite{bruggeman1935,landauer1952,zeng1988,rossi2009}.
In the EMT the inhomogeneous system is replaced by an effective homogeneous medium having the same, bulk,
electric properties. Let $\epsilon_M$, $\epsilon_I$ be the dielectric constants (at the probing frequency)
of \vo in the metallic and insulating phase respectively.
Then, the dielectric constant
of the effective medium, $\epsilon_{EMT}$, is given by the following equation:
\beq
  \frac{\eta_I(\epsilon_I - \epsilon_{EMT})}{ \epsilon_{EMT} + g(\epsilon_I - \epsilon_{EMT})} + \frac{\eta_M( \epsilon_M - \epsilon_{EMT}) }{ \epsilon_{EMT} + 
    g(\epsilon_M -  \epsilon_{EMT})} = 0.
 \label{eq:emt}
\enq
In Eq.~\ceq{eq:emt} $g$ is a factor that depends on the shape of the grain. 
Without loss of generality we set $g=1/3$.
Let $n$ and $k$ be the real and imaginary parts respectively of the index of refraction, so that for the effective medium $n+ik=\sqrt{\epsilon_{EMT}}$ and therefore
\beq
  R = \left| \frac{\cos \theta_0 - \sqrt{(n+ik)^2-\sin^2\theta_0} }{\cos \theta_0 + \sqrt{(n+ik)^2-\sin^2\theta_0}} \right|^2\;,
  \label{eq:remt}
\enq
where $\theta_0\approx 15^\circ$ corresponds to the probe incidence angle.
Given our experimental setup we can reliably obtain the imaginary part of the index of refraction 
by measuring the absorption. For this reason  we set the
value of the imaginary part of the complex index of refraction $k_M$, ($k_I$) for the metallic and (insulating) phase to the measured values, consistent with the values reported in the literature~\cite{Verleur1968a,KanaKana2011}, and then use Eq.~(\ref{eq:remt})
and  the measured value of $R_{M}$ ($R_{I}$) in the metallic (insulating) phase to fix the corresponding value of $n_M \; (n_I)$ (see Table \ref{table:parameters2}).

Using Eqs.~\ceq{eq:etaI}-\ceq{eq:remt} we can obtain the profile of $R(T)$ across the MIT for a given $P(T_c)$.
Assuming $P(T_c)$ to be a Gaussian distribution, by fitting the measured profile of $R(T)$ to the one
obtained using Eqs.~\ref{eq:etaI}-\ref{eq:remt} we can obtain the average value of the critical temperature $\langle T_c\rangle$
and its standard deviation $\sigma_{T_c}$. 
For \votio samples we find $\tav=314$~K, $\sigma_{T_c}=2.6$~K, 
for \vosap samples $\tav=340$~K, $\sigma_{T_c}=8.8$~K.
The solid lines in Fig.~\ref{fig:01} show the profiles of $R(T)$ obtained using Eqs.~\ceq{eq:etaI}-\ceq{eq:remt} 
and the above values for $\tav$ and $\sigma_{T_c}$.

The difference in the value of $T_c$ between \votio and \vosap
  samples can be attributed to the fact that \tio, having a rutile structure,
  might induce strains into the \vo film that should favor the metallic 
  phase of \vo. In general, strain effects are expected to play an important role
  in the physics of the MIT phase transition of \vo films.
  In our approach such effects enter indirectly, via the
  form of the probability distribution $P(T_c)$, and the value
  of the thermal conductivity of the interface between the \vo film
  and the substrate.

As we discuss in the following section,  
for our theoretical treatment of the recovery dynamics over long timescales of \vo films
the knowledge of $P(T_c)$, i.e., $\tav$ and $\sigma_{T_c}$,
is all that is needed. As mentioned before the fact that $\sigma_{T_c}$ is nonzero is due to
inhomogeneities, of different nature, present in the \vo film.
It is practically impossible to know the distribution in the films of the properties 
affecting $T_c$. However, it is interesting to consider the limit in which the 
grain size $D$ is the dominant property affecting $T_c$. The reason is that in this limit it is possible
to extract, using strong and fundamental arguments, the distribution, $P(D)$, for the grain size.
In particular, it is possible to obtain the average grain size, $\dav$, and its standard deviation, 
quantities that are of great practical interest. 
$\dav$ can be compared to estimates obtained using more direct
experimental techniques, such as XRD.
In the remainder of this section we use the experimental results for $R(T)$ across the MIT
to extract $\dav$ and its standard deviation.

Theoretical and experimental evidence\cite{granqvist1976} indicates that for thin films the distribution $P(D)$ 
of the grain size $D$ typically follows a logarithmic-normal distribution,
\beq
 P(D) = \frac{1}{\sqrt{2\pi} \sigma D} \exp {\left[ -\frac{\left[\ln {D/\hat D}  \right]^2}{2 \sigma^2} \right]}.
 \label{eq:PD}
\enq 
In Eq.~\ceq{eq:PD} $D$ is the effective diameter of a grain, $\hat D$ is the grain size (diameter)
such that $\ln\hat D=\langle\ln D\rangle$,
and $\sigma$ is the standard deviation of $\ln(D)$.

From general and fundamental arguments~\cite{fisher1982,challa1986,zhang2000,jiang2002} we have:
\beq
 T_c = T^{\rm (bulk)}_c \left(1 - \frac{1}{D/D_0} \right)\;,
 \label{eq:Tc}
\enq
where  $T^{\rm (bulk)}_c$ is the bulk transition temperature and $D_0$, equal to 2~nm in our case,
is the grain's diameter below which the grain is so small that is not possible to unambiguously identify its crystal structure.
We set $T_c^{\rm(bulk)}=355$~K, that is the temperature above which the \vosap samples is completely metallic.
This value is higher than the value of bulk \vo due to the strain experienced by the films \cite{brassardAPL05,alievPSS06}.
The relation between $P(D)$ and $P(T_c)$ is given by:
\beq
 P(T_c)=P(D(T_c))\frac{dD}{dT_c}.
 \label{eq:PTc}
\enq

\begin{figure}[!!!!h!!!!!bt]
    \includegraphics[width=8.5cm]{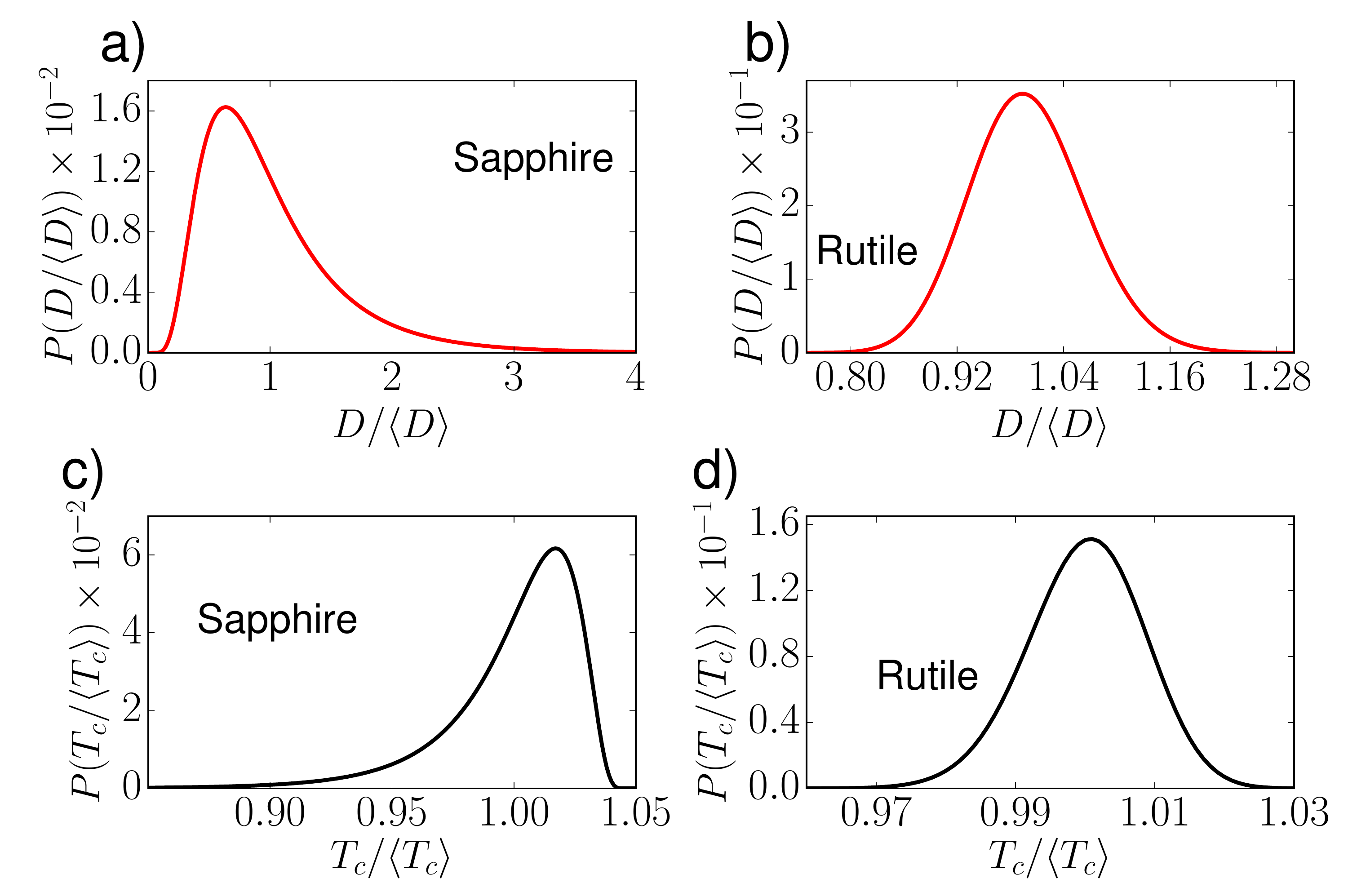}
    \caption{(a) and (b) show the grain size distributions normalized to the average grain size for sapphire ($\langle D \rangle = 64.7$ nm) and rutile ($\langle D \rangle = 17.4$ nm) substrate respectively. (c) and (d) show the critical temperature distribution normalized to the average critical temperature for sapphire ($\langle T_c \rangle = 340.1$ K) and rutile ($\langle T_c \rangle = 314.0$ K) respectively. The bulk critical temperature is taken to be $T_c^{(bulk)}=355$ K.
            } 
    \label{fig:02}
\end{figure}

\begin{figure}[!!!!h!!!!!bt]
    \includegraphics[width=8.5cm]{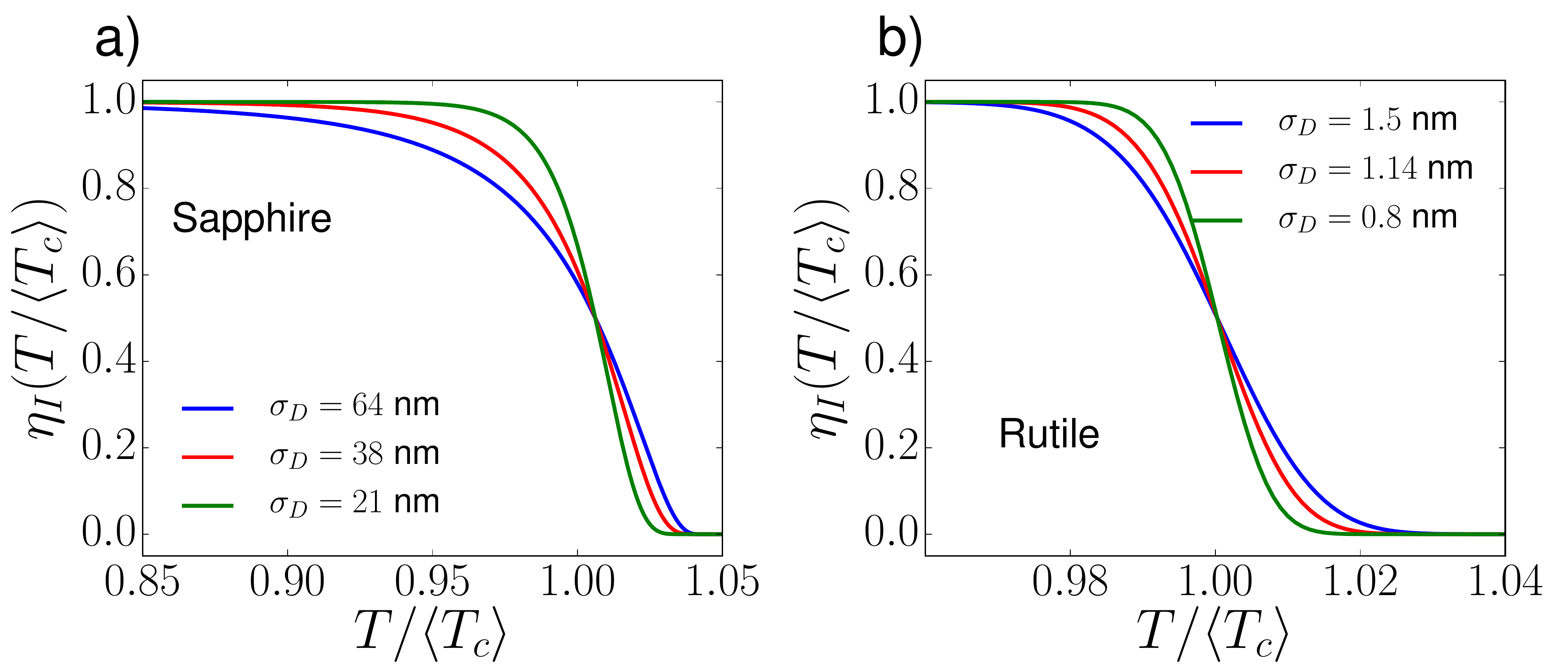}
    \caption{Evolution of the insulating partial volume $\eta_I$ across the thermally induced MIT for case of (a) sapphire and (b) rutile substrates. For rutile, $\langle T_c \rangle = 314.0$ K, and for sapphire $\langle T_c \rangle = 340.1$ K.
             }
   \label{fig:03}
\end{figure}

Using Eqs.~\ceq{eq:etaI}-\ceq{eq:PTc}, by fitting the measured profile of $R(T)$ across the MIT, we can 
obtain $P(D)$ and therefore $\dav$ and its standard deviation.
Figures~\ref{fig:02}~(a),~(b) show the profiles of $P(D)$ 
used to obtain the good theoretical fit to the evolution
of $R(T)$ shown in Fig.~\ref{fig:01}.
These profiles correspond to
$\dav=64.7$~nm  $\sigma_D=38.5$~nm for \vosap samples, and
$\dav=17.4$~nm  $\sigma_D=1.1$~nm \votio samples.
It is interesting to compare the values of $\dav$ obtained using this approach 
to the values estimated using XRD.
From XRD data \cite{radue2015} we estimated $\dav\approx 45$~nm for \vosap and $\dav\approx 13$~nm for \votio (see Table \ref{table:parameters2}).
These values are in remarkable semi-quantitative agreement with the values extracted from the profiles of $R(T)$
across the MIT suggesting that the assumption that the grain size is the dominant property affecting 
the local value of $T_c$ might be qualitatively correct. 
It is therefore interesting to obtain the profiles of $P(T_c)$ corresponding to the distributions of grain sizes
shown in Figs.~\ref{fig:02}~(a),~(b). Such profiles are shown in Figs.~\ref{fig:02}~(c),~(d).
The evolution of $\eta_I(T)$ across the MIT obtained using these profiles is shown in Fig.~\ref{fig:03}.
\begin{table}[ht]
\centering 
\begin{tabular}{l l l} 
\hline\hline
                        &  \votio            &  \vosap \\
\hline                  
 $\langle T_c \rangle$  &  $ 314.0$ K          &  $340.1$ K     \\
 $\sigma_{T_c}$         &  $ 2.6$   K          &  $8.8 K$      \\
 $\langle D \rangle$     &  $17.4$ nm           &  $64.7$ nm          \\
 $\langle D \rangle_{Exp} \cite{radue2015}$     &  $13$ nm           &  $45$ nm          \\
  $\sigma_D$            &  $ 1.1$ nm           &  $38.5$ nm     \\
 $n_{M}+ik_M$           &   $1.53 + i0.8$      &   $1.49 + i0.65$        \\  
 $R_{M}$                &  $ 0.14$             &  $0.11$       \\  
 $n_{I}+ik_I$           &   $3.03 + i0.57$     &   $2.60 + i0.60$        \\  
 $R_{I}$                &  $ 0.28$             &  $0.23$    \\  
 $\sigma_K$            & $1,100$ W/(K cm$^2$) & $13,000$ W/(K cm$^2$) \\ 
\hline\hline
\end{tabular}
\caption{ Comparative table between \votio, and \vosap sample parameters. } %
\label{table:parameters2} %
\end{table}
%

Our analysis  suggest that the $R(T)$ profiles could be
an indirect method to characterize the distribution of grain sizes in \vo films,
a very challenging quantity to obtain using direct imaging experiments.

\section{Theoretical modeling of the relaxation dynamics of the MIT}
\label{sec:th_model_dynamics}

\begin{figure}[th!]
\centering
\includegraphics[width=4cm]{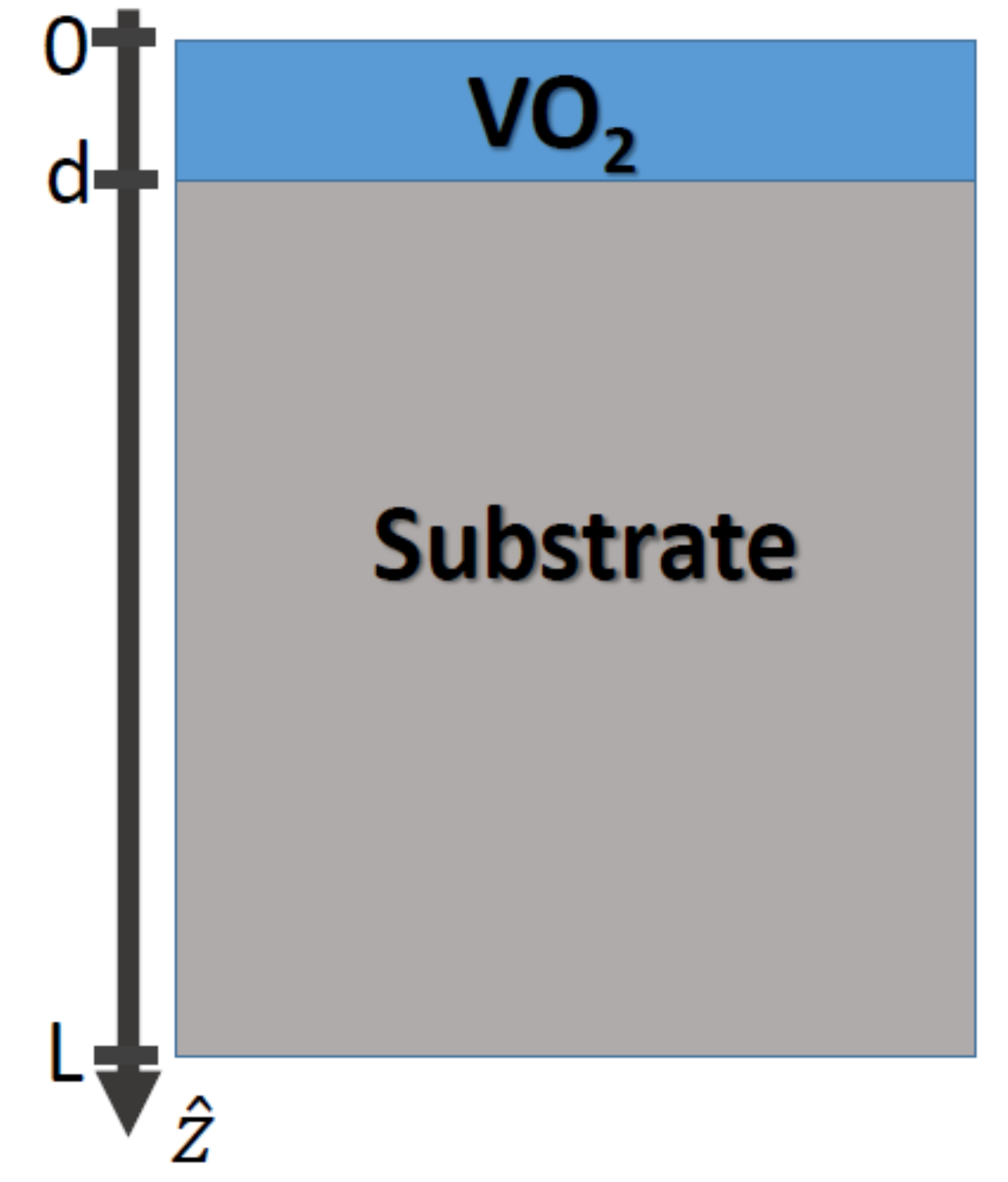}
\caption{Sketch of the heterostructure considered in this work. It is composed of a vanadium dioxide (VO$_2$) thin-film deposited on top of a substrate. The substrates considered in this work are titanium dioxide (TiO$_2$), and aluminum oxide (Al$_2$O$_3$). For VO$_2/$TiO$_2$ $d=110\;$nm while for VO$_2/$Al$_2$O$_3$ $d=80\;$nm. For both substrates, $L=0.5\;$ mm. }
\label{fig:structure}
\end{figure}

In our experiment the \vo films have a thickness $d$ equal to or smaller than $110$~nm (see Fig.~\ref{fig:structure}), which is comparable with the laser $1/e$ penetration depth $\delta\simeq 110-130$~nm~\cite{radue2015}. Thus,  
we can assume that the pump pulse heats the film uniformly throughout its thickness.
To describe the heat transfer process between the film and 
the substrate, we assume the temperature to be uniform throughout
the film for all times. Effectively, given these conditions,
the heat transfer problem becomes a one-dimensional problem,
and the equation for the rate of change of the heat ($Q$) in the film takes the form:
\begin{align}
\frac{dQ}{dt} =& A\times d\times \left( \right. \rho_I C_I \eta_I(T_f) + \rho_M C_M \eta_M(T_f)   \nonumber \\
               & +\left. L(T_f)P(T_f)\rho_{av} \right) \frac{\partial T_f}{\partial t}\;,
\label{eq:dQ}
\end{align}
where $T_f$ is the film temperature, $A$ is the area of the film,
$\rho_I$ ($\rho_M$) 
is the density in the insulating (metallic) phase, $\rho_{av}\df(\rho_I+\rho_M)/2$,
$C_{I}$ ($C_{M}$) 
is the heat capacity in the insulating (metallic) phase, 
$L$ is the specific heat and $P(T_f)dT_f$ is the fraction of the sample
undergoing the MIT in the time interval $dt$ during which the 
film temperature is in the interval $[T_f,T_f+dT_f]$. 
Here $P(T_f)$ is the distribution of critical temperatures due to the inhomogeneities
that we have obtained in the previous section.
Using Eq.~\ceq{eq:etaI} and the fact that $\eta_M=(1-\eta_I)$ we can rewrite Eq.~\ceq{eq:dQ}
in a form that more explicitly shows the effect due to the inhomogeneities, i.e. the fact
that the MIT is not characterized by a single $T_c$, but by a distribution of $T_c$'s:
\begin{align}
 \frac{dQ}{dt} = A\times d\times &\left[\rho_M C_M + (\rho_I C_I -\rho_M C_M)\int_{T_f}^\infty P(T_c)dT_c\right. \nonumber \\
                & + L(T_f)P(T_f)\rho_{av} \bigg] \frac{\partial T_f}{\partial t}\;.
 \label{eq:dQ2} 
\end{align}
Equation~\ceq{eq:dQ2} is the main result of our work: it allows to properly take into account
the effect of inhomogeneities on the long timescale dynamics across a first order phase transition.
The key quantity entering Eq.~\ceq{eq:dQ2} is the distribution $P(T_c)$ that, as we have shown in
the preceding section, can be obtained from the profile of $R(T)$ across the thermally activated
MIT. Our work is the first to combine the information from the thermally activated MIT
to obtain a physically accurate heat equation to describe the recovery dynamics after a photo-induced MIT.
%

%
For the latent heat we have \cite{fisher1982,challa1986,zhang2000}
\beq
 L = L^{\rm (bulk)} \frac{T_c}{T_c^{\rm (bulk)}}.
 \label{eq:L}
\enq
where $L^{\rm (bulk)}$ is the value of the specific heat for bulk \vo.
Given Eq.~\ceq{eq:Tc}, Eq.~\ceq{eq:L} implies
$L = L^{\rm (bulk)}(1 - D_0/D)$.

The rate of change of heat in the film given by Eq.~\ceq{eq:dQ} must be equal to
the heat current ($J_Q$) across the interface between the film and the substrate:
\beq
  J_Q = -\sigma_K A(T_f - T_s(d))
  \label{eq:JQ}
\enq
where $\sigma_K$ is the Kapitza constant characterizing the
thermal conductivity of the interfaces \cite{Kapitza1941,Pollack1941,stoner1993,haidan2013},
and $T_s(d)$ is the temperature of the substrate at the surface facing the \vo film.
Combining Eq.~\ceq{eq:dQ2} and Eq.~\ceq{eq:JQ},  
for $T_f$ we obtain the equation:
\begin{align}
 &\Big[\rho_M C_M + (\rho_I C_I -\rho_M C_M)\int_{T_f}^\infty P(T_c)dT_c \nonumber \\
               &  + L(T_f)P(T_f)\rho_{av} \Big] \frac{\partial T_f}{\partial t} 
 = -\frac{\sigma_K}{d}(T_f - T_s(d)).
\label{eq:Tf}
\end{align}
%
%
In Eq.~\ceq{eq:Tf} the only undetermined quantity is $\sigma_K$.
We fix $\sigma_K$ by fitting the theoretically obtained time evolution of $R(t)$
to the one measured experimentally, {\em for fixed experimental conditions} such
as the temperature of the substrate and the pump fluence. 
The robustness of the theory presented is evidenced by the fact that, the same
{\em fixed} value of $\sigma_K$ provides a good agreement between the theoretical
and the experimental results for a broad range of experimental conditions.

To completely define the problem we need to supplement Eq.~\ceq{eq:Tf} with proper boundary conditions.
The temperature distribution within the substrate, $T_s(z,t)$, satisfies the diffusion equation:
\beq
 \frac{\partial T_s(z,t)}{\partial t} = \frac{k_s}{C_{s} \rho_s} \frac{\partial^2 T_s(z,t)}{\partial z^2}
 \label{eq:Ts}
\enq
where $k_s$, $C_s$, $\rho_s$ are the thermal conductivity, heat capacity, and mass density, respectively, of the substrate.
The bottom of the substrate, for which $z=L$ (see Fig.~\ref{fig:structure}), is kept at a fixed temperature $T_s^{(B)}$.
At the film/substrate interface the heat transferred from the film must be equal to the heat current $k_s\partial T_s/\partial z|_{z=d}$.
We then have that the boundary conditions completing Eq.~\ceq{eq:Ts} are:
\begin{align}
 &T_s(z=L,t)=T_s^{(B)};\label{eq:bc1}\\
 &k_s \left. \frac{\partial T_s(z,t)}{\partial z}\right|_{z = d} = -\sigma_K (T_f(t) - T_s(z=d,t) ). \label{eq:bc2}
\end{align}

Equations \ceq{eq:L},\ceq{eq:Tf}-\ceq{eq:bc2}, combined with knowledge of the distribution $P(T_c)$
completely define the temperature evolution of the \vo film.
Notice that in these equations the only unknown parameter is the Kapitza constant $\sigma_K$. All the other quantities
are known from direct measurements. $P(T_c)$ is obtained from the profile of $R(T)$ across the MIT,
independently from the dynamics of $R$ after the photo-induced insulator-to-metal transition.
Also, the relation between the specific heat $L$ and $T$ is fixed by general and fundamental
results \cite{fisher1982,challa1986,zhang2000}.

While these equations can  in general be solved only numerically, some qualitative understanding of the decay time $\tau$ can be gained if we make some simplifications.  
Let 
${P}(T_c)=1/(\sqrt{2\pi}\sigma_{T_c}) \exp\{-(T_c - \langle T_c \rangle)^2/(2 \sigma^2_{T_c})\} $.
At a temperature $T$ the insulating volume fraction is given by $\eta_I(T) = \frac{1}{2}\left[ 1- {\rm erf}
 \left( (T-\langle T_c \rangle)/(\sqrt{2}\sigma_{T_c}) \right)\right]$. Then assuming that the pump pulse is strong enough to drive the entire film into a fully metallic state at $t=0$, the time-dependence of the insulating volume fraction can be approximated by a simple exponential form $\eta_I(t) = 1-A e^{-t/\tau}$.
In this case, an expression for the temperature can be obtained through the relationship $\eta_I(T(t))=\eta_I(t)$. 
Furthermore, assuming that the substrate temperature $T_s$ does not change with time,  and the latent heat $L$ to be temperature-independent, we can calculate the decay constant $\tau$:
\beq
\tau = C d \frac{\sigma_{T_c}}{\sigma_K} \frac{\left( C_M \rho_M + L \rho_{av}P(T_0) \right)}{T_0 - T_s}+\tau_0\;,
\label{eqn:approximation}
\enq
where the constants $C>0$, and $\tau_0$ can only be determined by solving the full system of equations~\ceq{eq:Tf}-\ceq{eq:bc2}

\begin{figure}[!!!!h!!!!!bt]
    \includegraphics[width=8.5cm]{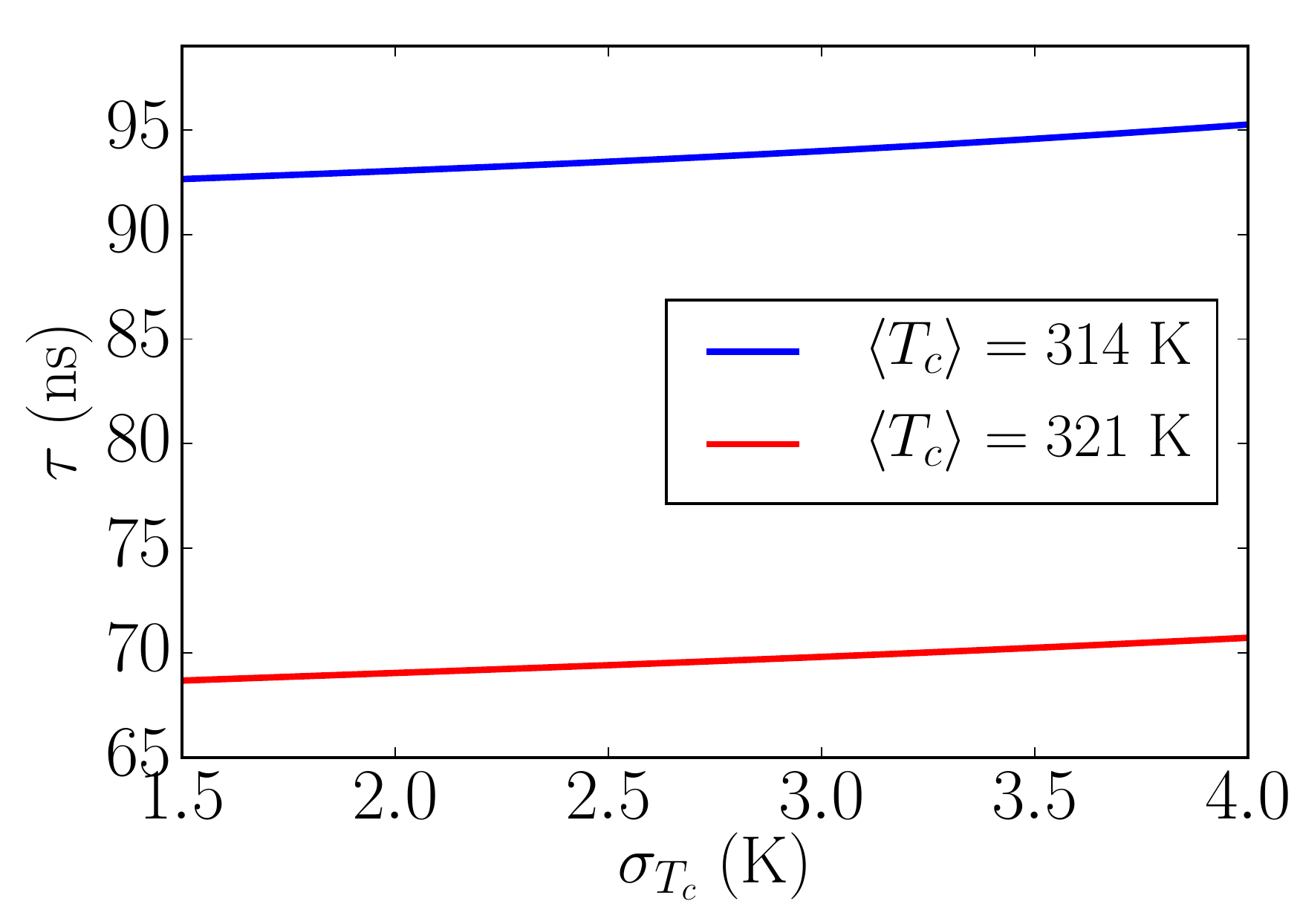}
    \caption{Full numerical calculation of the dependence of metal state decay constant $\tau$ on $\sigma_{T_c}$ for two different values of the sample average critical temperature $\langle T_c \rangle$, and $T_s(L)=280$~K. The initial temperature $T_0=360$~K is such that the sample is initially fully metallic, and $(T_0 - \langle T_c \rangle)/(\sqrt{2}\sigma_{T_c}) \approx 9$.}
    \label{fig:12}
\end{figure}

\begin{figure}[!!!!h!!!!!bt]
    \includegraphics[width=6.5cm]{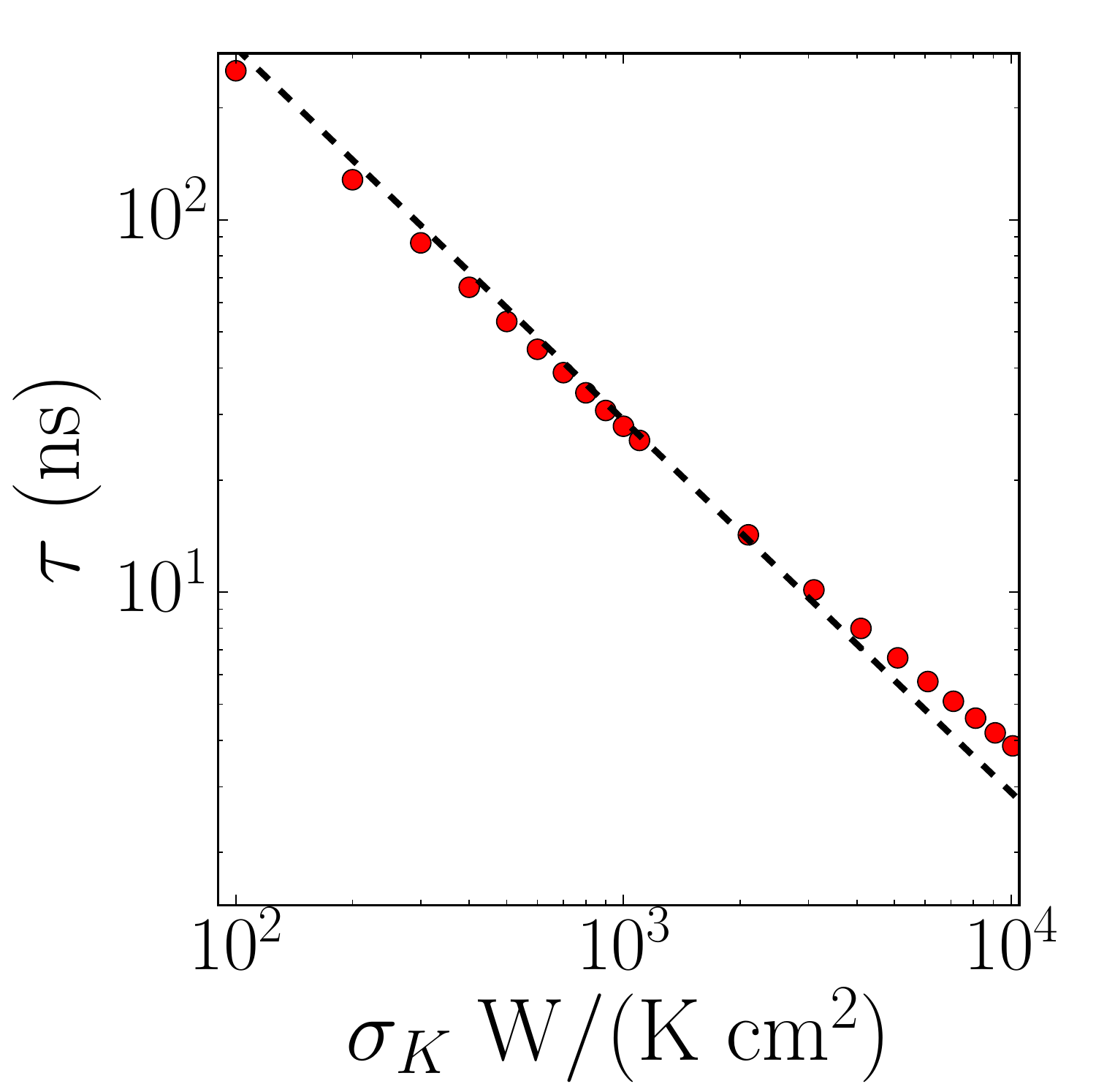}
    \caption{\vosap metal state decay time $\tau$ dependence on the Kapitza constant $\sigma_k$ for $\langle D \rangle=64.7$~nm, $\sigma_D=38.5$~nm, substrate temperature $T_s(L)=310$~K, and fluence $\phi = 8$~mJ/cm$^2$. The red dots correspond to numerical calculations, and the dashed line is given by $\tau \propto \sigma_K^{-1}$ }
    \label{fig:07b}
\end{figure}

It is interesting to note that despite its many limitations, Eq.(\ref{eqn:approximation}) captures many important qualitative traits of the actual relaxation process. For example, Figure~(\ref{fig:12}) shows a plot of the decay constant as a function of $\sigma_{T_c}$ obtained solving the full system of equations \ceq{eq:Tf}-\ceq{eq:bc1} for two different values of the average critical temperature, and same initial temperature $T_0=360$~K. It is easy to see that the decay time $\tau$ follows the linear trend predicted by Eq.~(\ref{eqn:approximation}) in the limit $(T_0 - \langle T_c \rangle)\gg \sqrt{2}\sigma_{T_c}$.
Similarly, an exact solution shows the inverse dependence of $\tau$ on the Kapitza constant, $\tau \propto \sigma_K^{-1}$, as shown in Figure~\ref{fig:07b}. 

The relation~\ceq{eqn:approximation} is another important result of our work, it shows how the characteristic time of the recovery dynamics is related
to the properties of the \vo films. In particular it shows the novel result that $\tau$ grows linearly with $\sigma_{T_c}$, the standard deviation of $P(T_c)$.
$\sigma_{T_c}$ can be reliably obtained from the profile of $R(T)$ across the MIT. 

\begin{table}[ht]
\centering 
\begin{tabular}{l c } 
\hline\hline  
VO$_2$ heat capacity insulating phase $C_I$  \cite{berglund1969} & 0.656 J/(g K)  \\ 
\hspace*{0.7cm} heat capacity metallic phase $C_M$    \cite{berglund1969} & 0.78 J/(g K)  \\
\hspace*{0.7cm} density insulating phase $\rho_I$     \cite{haidan2013}   & 4.57 g/cm$^3$  \\
\hspace*{0.7cm} density metallic phase $\rho_M$       \cite{haidan2013}   & 4.65 g/cm$^3$  \\
\hspace*{0.7cm} thermal conductivity insulating phase $\kappa_I$ \cite{oh2010}      & 3.5 W/(m K)  \\
\hspace*{0.7cm} thermal conductivity metallic phase $\kappa_M$   \cite{oh2010}      & 6 W/(m K)  \\
\hspace*{0.7cm} bulk latent heat $L^{(Bulk)}$ \cite{berglund1969}      & 51.8 J/g  \\
\\
TiO$_2$ heat capacity $C_{s}$   \cite{ligny2002}                              & 0.686 J/(g K)  \\ 
\hspace*{0.7cm} density $\rho_s$           \cite{haidan2013}          & 4.25 g/cm$^3$  \\
\hspace*{0.7cm} thermal conductivity $\kappa_s$ \cite{thurber1965}                      & 8 W/(m K)  \\
\\
Al$_2$O$_3$ heat capacity $C_{s}$      \cite{Pishchik2009}                       & 0.779 J/(g K)  \\ 
\hspace*{0.9cm} density $\rho_s$       \cite{Pishchik2009}                            & 3.98 g/cm$^3$  \\
\hspace*{0.9cm} thermal conductivity $\kappa_s$    \cite{Pishchik2009}                & 30 W/(m K)  \\
\votio absorption coefficient at 800 nm  $\alpha$ \cite{radue2015}            &  0.01 nm$^{-1}$           \\
\vosap absorption coefficient at 800 nm  $\alpha$ \cite{radue2015}         &  0.0076 nm$^{-1}$           \\
\hline\hline
\end{tabular}
\caption{Parameters of VO$_2$ and substrates.  } %
\label{table:parameters} %
\end{table}
\section{Effect of inhomogeneities on the relaxation dynamics of the photo-induced MIT}
\label{sec:results}

Using the theoretical approach described in Sec.~\ref{sec:th_model_dynamics} we can obtain
the time evolution of the optical reflectivity $R$ through the MIT, as well as explain the significant difference in relaxation timescales between the two \vo samples considered. In all the numerical calculations we assume $C_I$, $\rho_I$,  $C_M$, $\rho_M$ to be equal to the bulk values for 
insulating and metallic \vo, see Table~\ref{table:parameters}.

The initial film temperature is fixed by the pump fluence taking into account
the Gaussian profile of the pulse and the fact that some of the heat
is lost by the film during the time interval $[0,t_0]$ for which our analysis does not apply,
$t=0$ is time at which the pump pulse hits the \vo film
and $t_0=10$~ns for \votio films and $t_0=0.5$~ns for \vosap films.

\begin{figure}[!!!!h!!!!!bt]
    \includegraphics[width=8.0cm]{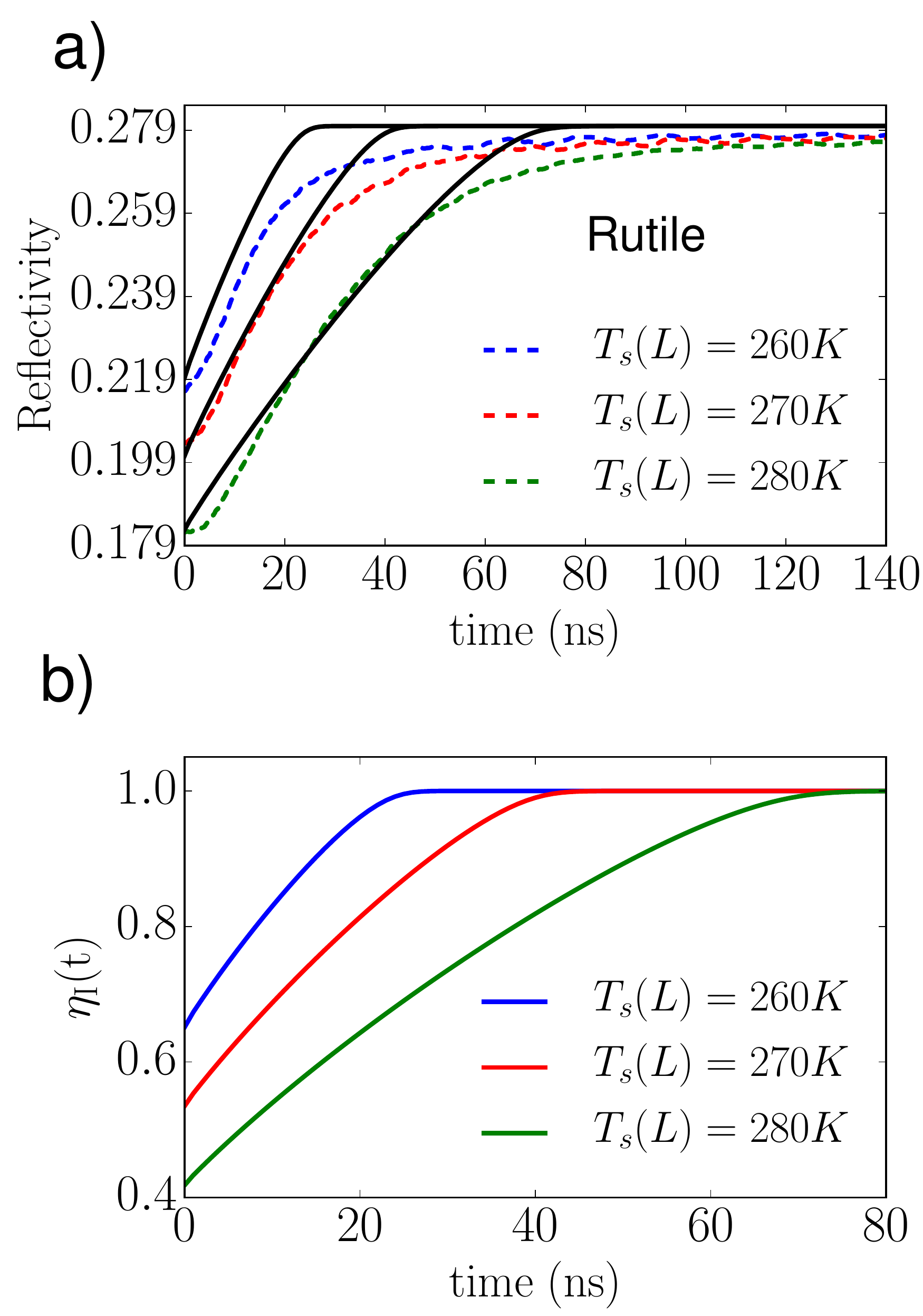}
    \caption{(a) Time evolution of reflectivity after the photo-induced MIT for \votio for three different $T_s(L)$ and $\phi = 9$ mJ/cm$^2$. The solid curves correspond to the theoretical results, and the dashed curves correspond to the experimental results. For the three theory curves we use $\sigma_K=1100$ W/(K cm$^2$). Panel (b) shows the corresponding insulating fraction time evolution.}
    \label{fig:05}
\end{figure}

\begin{figure}[!!!!h!!!!!bt]
    \includegraphics[width=8.5cm]{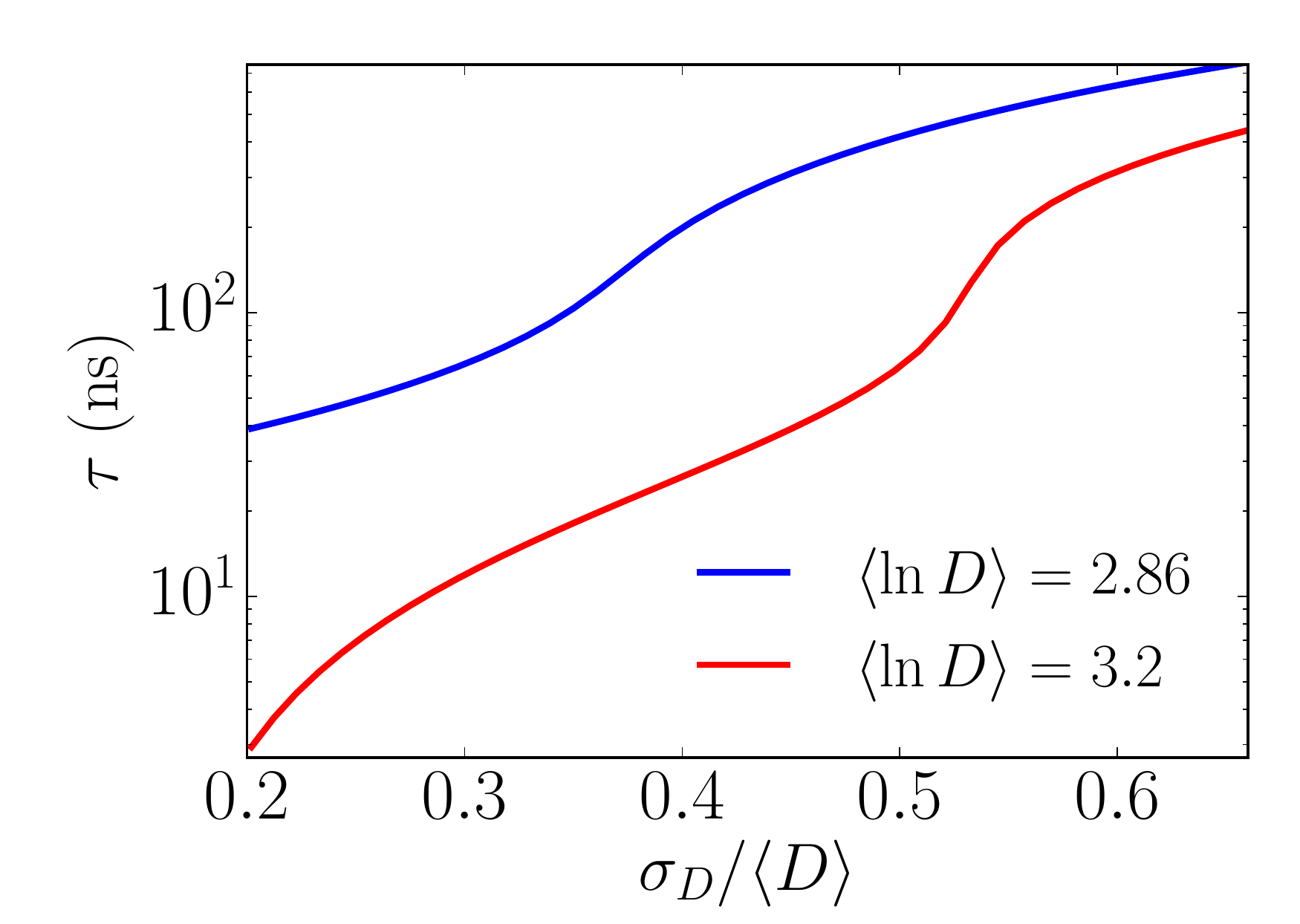}
    \caption{Dependence of the \votio metal state decay time constant $\tau$ on $\sigma_D$ for two values of $\langle \ln D \rangle$, as defined in Eq. (\ref{eq:PD}), Kapitza constant $\sigma_K=1100$ W/(K cm$^2$), substrate temperature $T_s(L)=280\;$ K, and initial fluence $\phi = 9$ mJ/cm$^2$. }
    \label{fig:06}
\end{figure}

\begin{figure}[!!!!h!!!!!bt]
    \includegraphics[width=8.5cm]{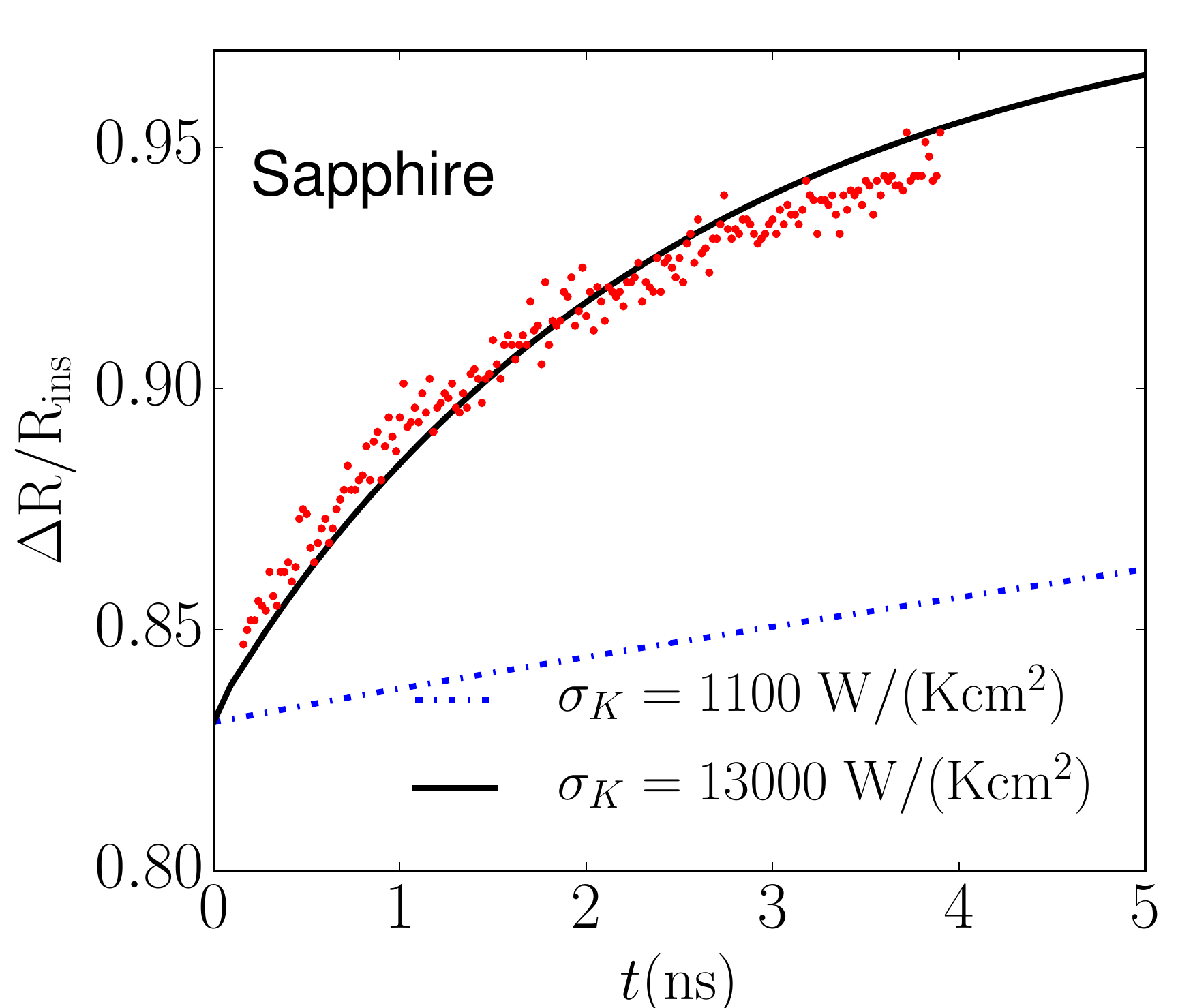}
    \caption{\vosap reflectivity time evolution after photo-induced MIT for $\phi = 7.5$ mJ/cm$^2$. The red dots correspond to the 
    experimental result. The dotted curve correspond to the theory with $\sigma_K=1100$ W/(K cm$^2$), and the solid curve 
    corresponds to $\sigma_K=13000$ W/(K cm$^2$).}
    \label{fig:07}
\end{figure}

As discussed in Sec.~\ref{sec:th_model_dynamics}, $\sigma_K$ is the only unknown parameter.
For the case of \votio samples, by fitting one of the curves for the dynamics of the reflectivity,
we find $\sigma_K=1100$ W/(K cm$^2$). 
We find that all experimental curves are well approximated assuming the same value for the Kapitza constant,
see Fig.~\ref{fig:05}~(a).
For the case of \vosap the characteristic timescale of the recovery is much shorter than for \votio samples.
As discussed in Sec.~\ref{sec:th_model_static} the two samples have very different inhomogeneities:
$\sigma_{T_c}$ is almost 4 times larger in \vosap than \votio. All other things being equal, 
Eq.~\ceq{eqn:approximation} implies, see Fig.~\ref{fig:06}, that $\tau$ should be larger in \vosap than in \votio, the opposite
of what is observed experimentally. We are then led to conclude that $\sigma_K$
in \vosap must be much higher than in \votio.
Figure~\ref{fig:07} shows the measured evolution of $R$ for the \vosap sample for a fixed
value of the fluence $\phi$ and substrate temperature, and the theoretical curves
for this case that we obtain using the distribution $P(T_c)$ obtained
for \vosap and two different values of $\sigma_K$. We see that by choosing for $\sigma_K$
the same value used for \votio, there is no agreement between theory and experiment.
By setting  $\sigma_K=13000$ W/(K cm$^2$) we obtain excellent agreement.
Indeed, all the experimental curves $R(t)$ for \vosap
are well approximated by the theoretical results assuming 
$\sigma_K=13,000$ W/(K cm$^2$).


\begin{figure}[!!!!h!!!!!bt]
    \includegraphics[width=8.5cm]{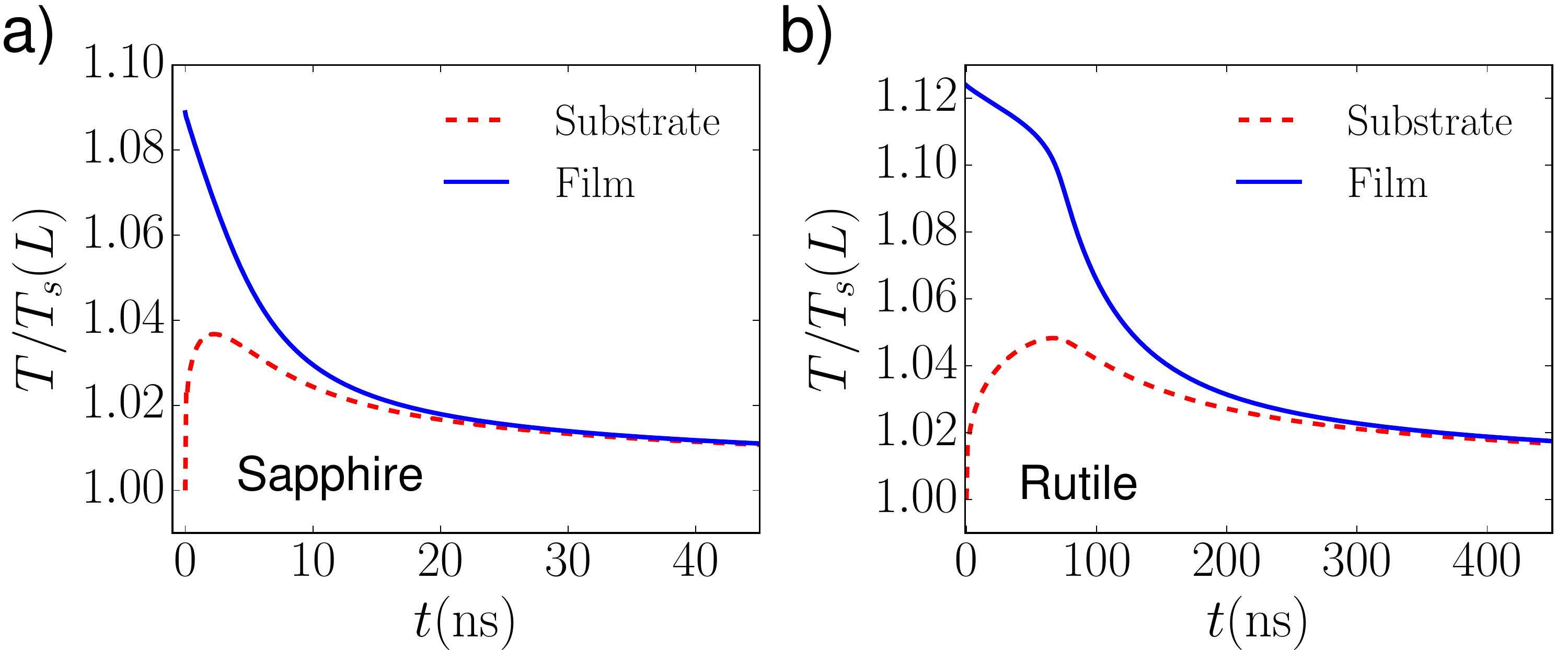}
    \caption{Film and substrate temperature time evolution. For sapphire (a), $T_s(L)=310$~K, and for rutile (b), $T_s(L)=280$~K. }
    \label{fig:08}
\end{figure}

Figure~\ref{fig:08} shows the time evolution of the \vo film and substrate temperatures (close to the interface) for the \vosap film, panel (a), and for the \votio film, panel (b), 
using the parameter values summarized in Table~\ref{table:parameters}. It helps to qualitatively understand the differences in the thermal relaxation between the two samples. Due to the lower values of the Kapitza constant, thermal energy stays more concentrated near the \vo-\tio interface, keeping the temperature of the \vo film above $T_c$ longer.

\begin{figure}[!!!!h!!!!!bt]
    \includegraphics[width=8.5cm]{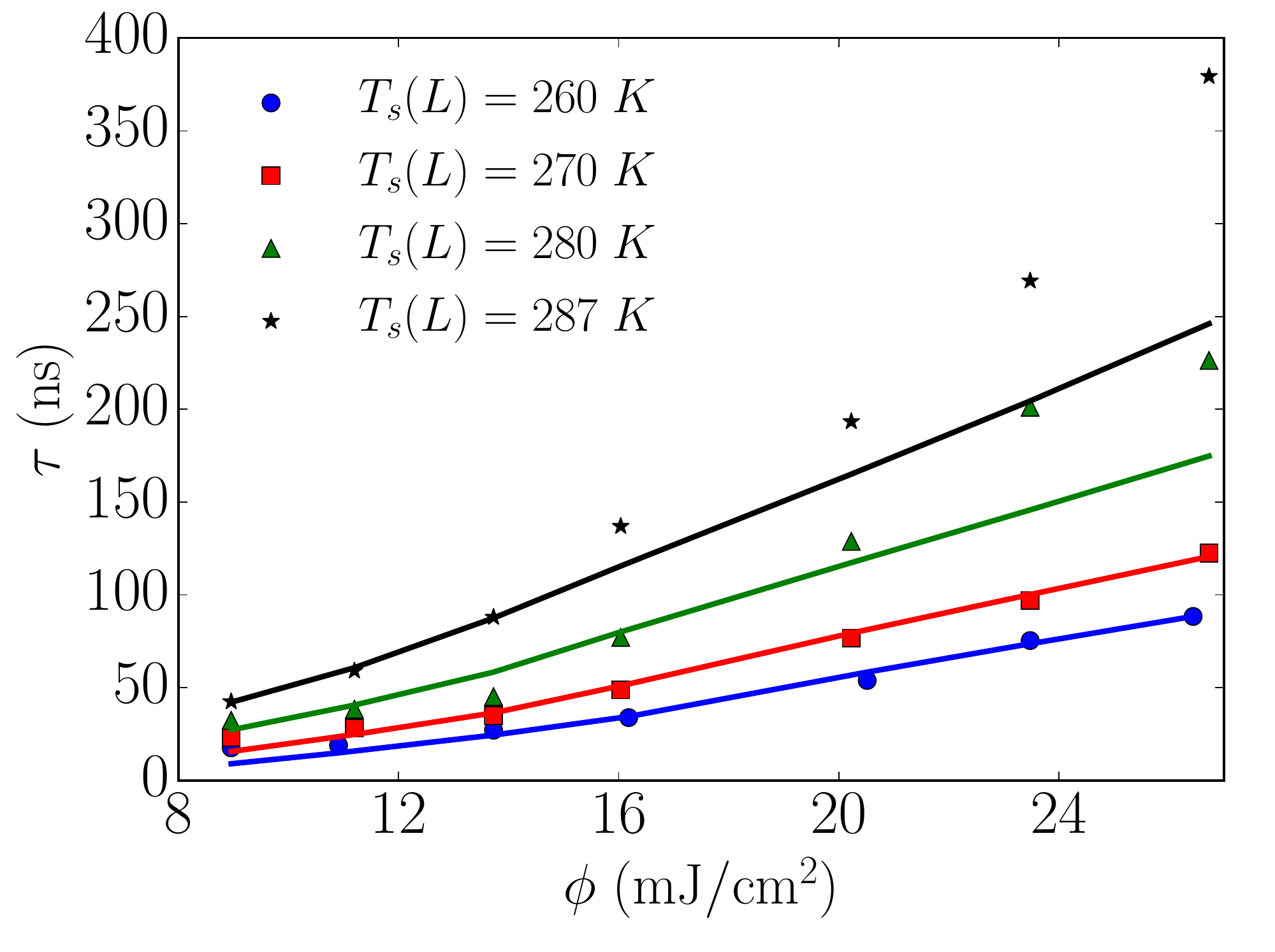}
    \caption{Dependence of metal state decay constant $\tau$ on fluence and substrate temperature for \votio.}
    \label{fig:10}
\end{figure}

To investigate the temperature dependence of the thermal relaxation we repeated the measurements while changing the base substrate temperature of the \votio sample. For these measurements the sample was placed inside a cryostat, and cooled down to temperatures $T_s(L)$ between $260$~K and $298$~K.
The results of these measurements, along with the theoretical calculations, are shown in Figure~\ref{fig:10}. We again observe a good semiquantitative
agreement between theoretical and experimental results. Also, note that the simple expression 
for the decay constant $\tau$ Eq. (\ref{eqn:approximation}) captures the overall 
decay rate drop at lower substrate temperatures $T_s(L)$.

We point out that all the theoretical curves are obtained
using the fixed set of parameters shown in Table~\ref{table:parameters}.
As mentioned above, the only unknown parameter that enters the theory is $\sigma_K$.
In the results presented above $\sigma_K$ was fixed to a single value for each film,
and this value was then used to obtain the results for a range of experimental
conditions with different substrate temperatures and pump fluences.
For example, Fig.~\ref{fig:09} shows an excellent agreement between the experimental measurements and theoretical calculations across the entire range of pump fluences,  limited on the lower end by our ability to reliably detect the variation in the probe reflectivity, and on the upper end by the damage threshold of our sample (pump fluence $>40 $mJ/cm$^2$).

%

\section{Conclusions}
\label{sec:conclusions}

In conclusion, we have presented a combined experimental and theoretical study of the long timescale
recovery dynamics of \vo films following an ultrafast photo-induced insulator-to-metal transition.
We have developed a theoretical approach that is able to properly
take into account the effect of inhomogeneities.
The main results of our work are:
(i)   The derivation of the generalized heat equation ~\ceq{eq:dQ2} that properly takes into account
      that during the recovery, due to the inhomogeneities, only fraction of the sample is undergoing
      the metal-to-insulator transition and correctly tracks the evolution in time of the metallic (insulating) phase;
(ii)  The clarification of the connection between the temperature dependent profile ($R(T$)) of the reflectivity
      across the thermally induced MIT and its dynamics after a photo-induced insulator-to-metal transition;
(iii)  The approximate relation, Eq.~\ceq{eqn:approximation}, between the characteristic time of the 
      recovery dynamics and the parameters of the film, in particular to the standard deviation of the
      distribution of critical temperatures as extracted from $R(T)$;
(iv) The ability of our theory to describe, using a fixed value of the Kapitza constant, the recovery
      dynamics for different values of the substrate temperature and pump fluence.
      By changing the pump fluence the characteristic time of the recovery can be changed, experimentally,
      by two orders of magnitude: our theory is able to account for such a change.

The theoretical approach that we present is general and can be used to describe
the dynamics (in the adiabatic limit) of inhomogeneous systems across a first order phase transition.
The approximate relation between the characteristic time $\tau$ and the parameters
of the system shows that $\tau$ is directly proportional to the width of the thermally
activated transition. This result allows to estimate the  recovery time of \vo films solely on the basis
of a measurement of $R(T)$ across the MIT. 

Assuming that variations of the size of the grains forming the films are the main source 
of inhomogeneities, using very general and fundamental relations between the grain
size and the grain's critical temperature, we have been able to obtain the distribution
of the grain sizes. In particular, we have been able to estimate the average grain's
size and its standard deviation. We find that the calculated average grain's size
is in remarkable semi-quantitative agreement with the one obtained from XRD measurements.
For systems in which inhomogeneities are mostly due to variations of the size $D$ of the grains,
our analysis provides a way to obtain the size distribution $P(D)$
from the temperature dependent profile of the reflectivity across the thermally induced MIT.
This could be very useful considering that $P(D)$ 
is a very challenging quantity to obtain via direct measurements.



\section{Acknowledgments}
This work was funded in part by NSF, DMR-1006013: Plasmon Resonances and Metal Insulator Transitions in Highly Correlated Thin Film Systems, Jeffress Trust Awards program in Interdisciplinary Research, ONR-N00014-13-1-0321, and the NASA Virginia Space Grant Consortium.
We also acknowledge support from the NRI/SRC sponsored ViNC center and the Commonwealth of Virginia through the Virginia Micro-Electronics Consortium (VMEC).

%


%

\end{document}